\documentclass[aps,preprint,amsmath,amssymb]{revtex4}
\usepackage{graphicx}
\usepackage{epstopdf}
\begin{document}

\title{Anomalous quartic $ZZ\gamma\gamma$ couplings at the CLIC}

\author{M. K\"{o}ksal}
\email[]{mkoksal@cumhuriyet.edu.tr} \affiliation{Department of
Physics, Cumhuriyet University, 58140, Sivas, Turkey}

\begin{abstract}
We study the sensitivity to anomalous quartic $ZZ\gamma\gamma$ couplings through the processes $e^{+}e^{-}\rightarrow Z\, Z\, \gamma$,
$e^{+}e^{-} \rightarrow e^{+}\gamma^{*} e^{-} \rightarrow e^{+} Z\, Z\, e^{-}$ and $e^{+}e^{-} \rightarrow e^{+}\gamma^{*} \gamma^{*} e^{-} \rightarrow e^{+}\, Z\, Z\, e^{-}$  at the CLIC. We find $95\%$ confidence level bounds on these coupling parameters defining by the dimension-six operators. The best bounds on the anomalous $ZZ\gamma\gamma$ couplings among the three processes are obtained from $e^{+}e^{-}\rightarrow Z\, Z\, \gamma$ at a center of mass energy of $3$ TeV and an integrated luminosity of $590$ fb$^{-1}$. We show that the best bounds obtained on both the anomalous coupling parameters are of the order of $10^{-8}$ GeV$^{-2}$, significantly improving the current bounds.
\end{abstract}

\maketitle

\section{Introduction}

In the context of the Standard Model (SM), self-interactions of gauge bosons are exactly determined by the $SU(2)_{L}\otimes U(1)_{Y}$ gauge symmetry.
For this reason, the measurement of these couplings plays an important role in finding out the gauge structure of the SM. Any deviation of the triple and quartic couplings of the gauge bosons from the SM expectations would indicate the existence of new physics. Investigation
of new physics through effective Lagrangian method is a well known approach. Such an approach is described by high-dimensional operators which give rise to genuine quartic gauge couplings. These effective operators also do not cause new trilinear vertices. Therefore, genuine quartic gauge couplings can be studied independently from trilinear couplings. Imposing custodial $SU(2)$ and local $U(1)$ symmetry, and if we
restrict ourselves to charge conjugation and parity conserving interactions, the dimension 6 effective Lagrangian for the $ZZ\gamma\gamma$ couplings is defined by the operators \cite{lag2,lag3}

\begin{eqnarray}
L=L_{0}+L_{c},
\end{eqnarray}

\begin{eqnarray}
L_{0}=\frac{-\pi\alpha}{4}\frac{a_{0}}{\Lambda^{2}}F_{\mu \nu}F^{\mu \nu} W_{ \alpha}^{(i)} W^{(i) \alpha},
\end{eqnarray}

\begin{eqnarray}
L_{c}=\frac{-\pi\alpha}{4}\frac{a_{c}}{\Lambda^{2}}F_{\mu \alpha}F^{\mu \beta} W^{(i)\alpha} W_{\beta}^{(i)}
\end{eqnarray}
where $W^{(i)}$ is the $SU(2)_{weak}$ triplet, $\Lambda$ stands for the new physics energy scale, $F_{\mu \nu}$ is the photon field strength tensor, and $a_{0}$ and $a_{c}$ are the dimensionless anomalous coupling constants.

The $Z^{\mu} Z^{\nu} \gamma^{\alpha} \gamma^{\beta}$ vertex functions produced by dimension 6 effective quartic Lagrangians are given by

\begin{eqnarray}
i\frac{2\pi\alpha}{\textmd{cos}^{2}\theta_{W}\Lambda^{2}}a_{0}g_{\mu\nu}[g_{\alpha\beta}(p_{1}.p_{2})-p_{2\alpha}p_{1\beta}],
\end{eqnarray}

\begin{eqnarray}
&&i\frac{\pi\alpha}{2 \textmd{cos}^{2}\theta_{W}\Lambda^{2}}a_{c}[(p_{1}.p_{2})(g_{\mu\alpha}g_{\nu\beta}+g_{\mu\beta}g_{\alpha\nu})+g_{\alpha\beta}(p_{1\mu}p_{2\nu}+p_{2\mu}p_{1\nu}) \nonumber \\
&&-p_{1\beta}(g_{\alpha\mu}p_{2\nu}+g_{\alpha\nu}p_{2\mu})-p_{2\alpha}(g_{\beta\mu}p_{1\nu}+g_{\beta\nu}p_{1\mu})]
\end{eqnarray}
where $p_{1}$ and $p_{1}$ are the momenta of photons.

The experimental bounds on the anomalous coupling parameters $\frac{a_{0}}{\Lambda^{2}}$ and $\frac{a_{c}}{\Lambda^{2}}$ have been obtained at the LEP by the L3 collaboration through the study of the process $e^{+}e^{-}\rightarrow Z \gamma \gamma \rightarrow q \bar{q} \gamma \gamma$, and by the OPAL collaboration from the combination of the processes $e^{+}e^{-} \rightarrow Z \gamma \gamma \rightarrow q \bar{q} \gamma \gamma$ and $e^{+}e^{-} \rightarrow Z \gamma \gamma \rightarrow\nu \bar{\nu} \gamma \gamma$. The bounds on quartic anomalous $ZZ\gamma\gamma$ couplings from the L3 collaboration are \cite{L3}:

\begin{eqnarray}
 -0.02\,  \textmd{GeV}^{-2}<\frac{a_{0}}{\Lambda^{2}}<0.03\, \textmd{GeV}^{-2},
\end{eqnarray}
\begin{eqnarray}
 -0.07\,  \textmd{GeV}^{-2}<\frac{a_{c}}{\Lambda^{2}}<0.05\, \textmd{GeV}^{-2}.
\end{eqnarray}

 More restrictive bounds are obtained by the OPAL collaboration \cite{OPAL}:

\begin{eqnarray}
-0.007\, \textmd{GeV}^{-2}<\frac{a_{0}}{\Lambda^{2}}<0.023\, \textmd{GeV}^{-2},
\end{eqnarray}
\begin{eqnarray}
-0.029\, \textmd{GeV}^{-2}<\frac{a_{c}}{\Lambda^{2}}<0.029\, \textmd{GeV}^{-2}.
\end{eqnarray}
All bounds are given at $95\%$ C. L..

Up to now, the anomalous quartic $ZZ\gamma\gamma$ couplings at the linear $e^{+}e^{-}$ colliders and its $e \gamma$ and $\gamma\gamma$ options
have been investigated via  the processes $e^{+}e^{-}\rightarrow Z \gamma \gamma$ \cite{stir,rodri},
$e^{+}e^{-}\rightarrow Z Z \gamma$ \cite{bel}, $e^{+}e^{-}\rightarrow q \bar{q} \gamma \gamma$ \cite{mon}, $e \gamma\rightarrow Z Z e$ \cite{sok},
$e\gamma\rightarrow Z \gamma e$ \cite{sato}, and $\gamma\gamma\rightarrow WWZ$ \cite{mem}. These vertices have  also been examined at hadron colliders through the processes $p\,\bar{p}\rightarrow Z \gamma \gamma $ \cite{der}, $p\, p\, (\bar{p})\rightarrow \gamma \gamma \ell \ell$ \cite{ebol},
$pp\rightarrow p \gamma^{*} \gamma^{*} p \rightarrow p Z Z p$ \cite{roy,had,gup,piot}, $pp\rightarrow p \gamma^{*} p \rightarrow p Z Z q X$ \cite{senol}, $pp\rightarrow p \gamma^{*} p \rightarrow p \gamma Z q X$ \cite{inanc}, $pp\rightarrow qq \gamma \ell \ell$ \cite{had3}.

\section{EQUIVALENT PHOTON APPROXIMATION AT THE LINEAR COLLIDER}

Thanks to the high center-of-mass energy and luminosity, the LHC is expected to answer some of the fundamental open questions in particle physics. However, for high precision measurements, a TeV scale $e^{+}e^{-}$ linear collider with extremely high luminosity and clean experimental environments must be built to complement the LHC. One of the most popularly envisaged linear collider is the Compact
Linear Collider (CLIC) \cite{17,18}. The CLIC has been planned to operate at three different energy stages, with the basic parameters of these stages given in Table I.
The other well-known applications of linear colliders are to investigate new physics beyond the SM with the aid of $e \gamma^{*}$ and $\gamma^{*}\gamma^{*}$ reactions. An almost real $\gamma^{*}$ photon emitted from either of the incoming leptons can interact with the other lepton, and the process $e^{+}e^{-}\rightarrow e^{+}\gamma^{*} e^{-}\rightarrow e^{+}\,Z\,Z\,e^{-}$ can take place, as shown in Fig. $1$. In addition, the photons emitted from both leptons can collide with each other, and the process $e^{+}e^{-} \rightarrow e^{+}\gamma^{*} \gamma^{*} e^{-} \rightarrow e^{+}\,Z\, Z\,e^{-}$ can take place, as depicted by Fig. $2$. Photons emitted from leptons are described by equivalent photon approximation (EPA) \cite{19,20,21,ter,bro}.
In the framework of EPA, since the emitted almost real photons have a low virtuality, they are assumed to be on mass shell.
Many examples of investigation of possible new physics beyond the SM through photon-induced reactions using EPA are available in the literature \cite{22,23,25,27,28,288,29,30,31,32,33,34,35,36,37,38,39,399,40,41,42,43}.

The fundamental aim of the study presented here is to investigate the physics potential of the CLIC
in probing anomalous quartic $ZZ\gamma\gamma$ couplings via the reactions $e^{+}e^{-}\rightarrow Z\, Z\, \gamma$,
$e^{+}e^{-} \rightarrow e^{+}\gamma^{*} e^{-} \rightarrow e^{+} Z\, Z\, e^{-}$, and $e^{+}e^{-} \rightarrow e^{+}\gamma^{*} \gamma^{*} e^{-} \rightarrow e^{+}\, Z\, Z\, e^{-}$.

\section{CROSS SECTIONS}

The Tree-level Feynman diagrams for the processes $e^{+}e^{-}\rightarrow Z\, Z\, \gamma$, and the subprocesses $e^{-}\gamma^{*} \rightarrow Z\,Z\,e^{-}$, $\gamma^{*} \gamma^{*}\rightarrow Z\,Z$ are shown in Figs. $3$-$5$. In the presence of the effective Lagrangians in equations $(2)$ and $(3)$, the processes $e^{+}e^{-}\rightarrow Z\, Z\, \gamma$ and the subprocesses $e^{-}\gamma^{*} \rightarrow Z\,Z\,e^{-}$ have only four Feynman diagrams.
As shown in Figs. $3$-$4$, while the top-left diagrams show the contribution coming from $ZZ\gamma\gamma$ coupling, the other diagrams originate from SM electroweak processes. As seen from Fig. $5$, the subprocess $\gamma^{*} \gamma^{*}\rightarrow Z\,Z$ has only a Feynman diagram which consists of new physics.
In this paper, we have used the COMPHEP-4.5.1 program \cite{comp} in order to obtain numerical estimates of the cross
sections of the processes under study. In our work, only one of the anomalous coupling parameters $\frac{a_{0}}{\Lambda^{2}}$ and $\frac{a_{c}}{\Lambda^{2}}$ is presumed to deviate from the SM at a time. The total cross sections for the processes $e^{+}e^{-}\rightarrow Z\, Z\, \gamma$, $e^{+}e^{-} \rightarrow e^{+}\gamma^{*} e^{-} \rightarrow e^{+} Z\, Z\, e^{-}$ as functions of anomalous $\frac{a_{0}}{\Lambda^{2}}$ and $\frac{a_{c}}{\Lambda^{2}}$ couplings are plotted in Figs. $6$-$8$.  As the value of the total cross section involving $\frac{a_{0}}{\Lambda^{2}}$ coupling is greater than the value of the $\frac{a_{c}}{\Lambda^{2}}$, we
anticipate that the bounds on the $\frac{a_{0}}{\Lambda^{2}}$ parameter will be more stringent than the bounds on $\frac{a_{c}}{\Lambda^{2}}$.

\section{SENSITIVITY TO THE ANOMALOUS COUPLINGS}

The number of expected events in the SM for the processes $e^{+}e^{-}\rightarrow Z\, Z\, \gamma$ and $e^{+}e^{-} \rightarrow e^{+}\gamma^{*} e^{-} \rightarrow e^{+} Z\, Z\, e^{-}$
are given by

\begin{eqnarray}
N_{SM}= L_{int} \times \sigma_{SM} \times BR(Z\rightarrow \ell \bar{\ell})^{2}\,\,\,\,\,\,\,\,\,\,\,\, (\ell=e,\mu)
\end{eqnarray}
where $L_{int}$ is the integrated luminosity and $\sigma_{SM}$ is the SM cross section. In this study we restrict to the decays of the Z bosons into $e^{-}$-$e^{+}$ and $\mu^{-}$-$\mu^{+}$, which represent easily detectable and background-free channels.
The branching ratio of the Z boson pairs in the final states of both processes is $BR(Z\rightarrow \ell \bar{\ell})^{2}=4.52\times 10^{-3}$.
The SM background cross sections for the processes $e^{+}e^{-}\rightarrow Z\, Z\, \gamma$ and
$e^{+}e^{-} \rightarrow e^{+}\gamma^{*} e^{-} \rightarrow e^{+} Z\, Z\, e^{-}$ are given in Table II. Here, we impose the acceptance cuts on the pseudorapidities  $|\eta^{e,\,\gamma}|<2.5$ and the transverse momentums $p_{T}^{\,\gamma}>20 \:$ GeV, $p_{T}^{\,e}>25 \:$ GeV  for the electron and the photon in the final state of both processes.

On the other hand, the SM cross section of the process $e^{+}e^{-} \rightarrow e^{+}\gamma^{*} \gamma^{*} e^{-} \rightarrow e^{+}\, Z\, Z\, e^{-}$ is quite small, because the subprocess $\gamma^{*} \gamma^{*} \rightarrow Z\,Z$ is not allowed at the tree level. It is only allowed at loop level and can be neglected \cite{roy,had,gup,piot}. Therefore, the observation of a few events at the end of such a process would be an important sign of new physics beyond the SM.

In our study, we use two different analyses to evaluate the sensitivity to the anomalous $ZZ\gamma\gamma$ couplings. First, we employ a simple one-parameter $\chi^{2}$ test when the number of SM events is greater than $10$. This analysis only applies in the case of the process $e^{+}e^{-}\rightarrow Z\, Z\, \gamma$ with center-of-mass energy of $\sqrt{s}=0.5$ TeV and integrated luminosity of $230$ fb$^{-1}$ since the SM event number is equal to $17$. The $\chi^{2}$ is defined as follows

\begin{eqnarray}
\chi^{2}=\left(\frac{\sigma_{SM}-\sigma_{NP}}{\sigma_{SM}\delta_{stat}}\right)^{2}
\end{eqnarray}
where $\sigma_{NP}$ is the total cross section in the presence of anomalous gauge couplings, $\delta_{stat}=\frac{1}{\sqrt{N}}$ is the statistical
error and here $N$ is the number of events.

On the other hand, for all other processes in the
study of the anomalous couplings we use a Poisson distribution, due to the number of SM events
being fewer than $10$. Sensitivity bounds are calculated by presuming the number of observed events to be equal to the SM prediction, i.e., $N_{obs}=L_{int} \times \sigma_{SM} \times BR(Z\rightarrow \ell \bar{\ell})^{2}$. The upper limits of
the number of events $N_{up}$ at the $95\%$ C.L. can be obtained as follows \cite{had,piot}

\begin{eqnarray}
\sum_{k=0}^{N_{obs}}P_{Poisson}(N_{up};k)=0.05.
\end{eqnarray}
The value of upper limits $N_{up}$ can be determined with respect to the value of the number of observed events \cite{poss}.  The number of observed events $N_{obs}$ and corresponding values for the upper limits $N_{up}$ at $95\%$ C.L. by using the Poisson distribution for the processes $e^{+}e^{-}\rightarrow Z\, Z\, \gamma$ and $e^{+}e^{-} \rightarrow e^{+}\gamma^{*} \gamma^{*} e^{-} \rightarrow e^{+}\, Z\, Z\, e^{-}$ with different values of luminosity and center-of-mass energy are given in Tables III-IV. Here, the calculated values for $N_{obs}$ are rounded to the nearest integer. For example, the number of observed events for the process $e \gamma\rightarrow Z\, Z\, e$ with $\sqrt{s}=1.5$ TeV is obtained as $0.1$ and $0.3$ for $L_{int}=100$ and $200$ fb$^{-1}$, respectively. Both values of the number of observed events have been rounded to $0$. For the process $e^{+}e^{-} \rightarrow e^{+}\gamma^{*} \gamma^{*} e^{-} \rightarrow e^{+}\, Z\, Z\, e^{-}$, for all center-of-mass energies and luminosities, the values of the upper limits is always $3$, since the number of observed events is
equal to $0$.

The upper bounds of the number of events $N_{up}$ at the $95\%$ C.L. can be transformed into bounds on the anomalous couplings $\frac{a_{0}}{\Lambda^{2}}$ and $\frac{a_{c}}{\Lambda^{2}}$. In Tables V-VII, we give the obtained one-dimensional bounds on anomalous couplings $\frac{a_{0}}{\Lambda^{2}}$ and $\frac{a_{c}}{\Lambda^{2}}$ at $95\%$ C.L. for the processes $e^{+}e^{-}\rightarrow Z\, Z\, \gamma$,
$e^{+}e^{-} \rightarrow e^{+}\gamma^{*} e^{-} \rightarrow e^{+} Z\, Z\, e^{-}$, and $e^{+}e^{-} \rightarrow e^{+}\gamma^{*} \gamma^{*} e^{-} \rightarrow e^{+}\, Z\, Z\, e^{-}$ at selected values of integrated luminosity and center-of-mass energy.
As can be seen in Tables V-VII, in the case $\sqrt{s}=0.5$ TeV and $L_{int}=10$ fb$^{-1}$ of data, the bounds on $\frac{a_{0}}{\Lambda^{2}}$ would be $[-3.24\times 10^{-4};\, 3.12\times 10^{-4}]$ GeV$^{-2}$ for $e^{+}e^{-}\rightarrow Z\, Z\, \gamma$, $[-3.30\times 10^{-4};\, 3.29\times 10^{-4}]$ GeV$^{-2}$ for $e^{+}e^{-} \rightarrow e^{+}\gamma^{*} e^{-} \rightarrow e^{+} Z\, Z\, e^{-}$, $[-1.00\times 10^{-4};\, 1.00\times 10^{-4}]$ GeV$^{-2}$ for $e^{+}e^{-} \rightarrow e^{+}\gamma^{*} \gamma^{*} e^{-} \rightarrow e^{+}\, Z\, Z\, e^{-}$ and $\frac{a_{c}}{\Lambda^{2}}$ would be $[-6.19\times 10^{-4};\, 5.70\times 10^{-4}]$ GeV$^{-2}$ for $e^{+}e^{-}\rightarrow Z\, Z\, \gamma$, $[-1.10\times 10^{-4};\, 1.09\times 10^{-3}]$ GeV$^{-2}$ for $e^{+}e^{-} \rightarrow e^{+}\gamma^{*} e^{-} \rightarrow e^{+} Z\, Z\, e^{-}$, $[-3.41\times 10^{-4};\, 3.41\times 10^{-4}]$ GeV$^{-2}$ for $e^{+}e^{-} \rightarrow e^{+}\gamma^{*} \gamma^{*} e^{-} \rightarrow e^{+}\, Z\, Z\, e^{-}$. We can easily understand that these bounds are more restrictive than the best limits obtained from OPAL collaboration.
In addition, in the case of the CLIC operates the maximum energy and luminosity, the bounds on $\frac{a_{0}}{\Lambda^{2}}$ would be $[-1.76\times 10^{-8};\, 1.75\times 10^{-8}]$ GeV$^{-2}$ for $e^{+}e^{-}\rightarrow Z\, Z\, \gamma$, $[-1.18\times 10^{-7};\, 1.18\times 10^{-7}]$ GeV$^{-2}$ for $e^{+}e^{-} \rightarrow e^{+}\gamma^{*} e^{-} \rightarrow e^{+} Z\, Z\, e^{-}$, $[-4.79\times 10^{-8};\, 4.79\times 10^{-8}]$ GeV$^{-2}$ for $e^{+}e^{-} \rightarrow e^{+}\gamma^{*} \gamma^{*} e^{-} \rightarrow e^{+}\, Z\, Z\, e^{-}$ and $\frac{a_{c}}{\Lambda^{2}}$ would be $[-3.06\times 10^{-8};\, 3.04\times 10^{-8}]$ GeV$^{-2}$ for $e^{+}e^{-}\rightarrow Z\, Z\, \gamma$, $[-4.55\times 10^{-7};\, 4.51\times 10^{-7}]$ GeV$^{-2}$ for $e^{+}e^{-} \rightarrow e^{+}\gamma^{*} e^{-} \rightarrow e^{+} Z\, Z\, e^{-}$, $[-1.79\times 10^{-7};\, 1.79\times 10^{-7}]$ GeV$^{-2}$ for $e^{+}e^{-} \rightarrow e^{+}\gamma^{*} \gamma^{*} e^{-} \rightarrow e^{+}\, Z\, Z\, e^{-}$. Hence, the best bounds for $\frac{a_{0}}{\Lambda^{2}}$ and $\frac{a_{c}}{\Lambda^{2}}$ improve the sensitivity by up to a factor of $10^{5}$ and $10^{6}$ with respect to current experimental bounds, respectively. Principally, it can be seen from a comparison of Tables V-VII that the sensitivity to anomalous couplings  rapidly increases when the center-of-mass energy and integrated luminosity of the processes increase.

\section{Conclusions}

Since the linear $e^{+}e^{-}$ colliders and its $e \gamma^{*}$ and $\gamma^{*} \gamma^{*}$ options have very clean experimental conditions and are mostly free from QCD backgrounds, measurements of the anomalous quartic gauge boson couplings with higher precision with respect to LHC can be obtained.
The anomalous quartic gauge couplings $\frac{a_{0}}{\Lambda^{2}}$ and $\frac{a_{c}}{\Lambda^{2}}$ are described by dimension $6$ effective quartic Lagrangian. The total cross sections containing these couplings have a stronger energy dependence than the pure SM processes. Therefore, having a high center-of-mass energy, the linear collider is extremely important  for the determination of the anomalous quartic gauge couplings. For these reasons, we study the processes $e^{+}e^{-}\rightarrow Z\, Z\, \gamma$,
$e^{+}e^{-} \rightarrow e^{+}\gamma^{*} e^{-} \rightarrow e^{+} Z\, Z\, e^{-}$, and $e^{+}e^{-} \rightarrow e^{+}\gamma^{*} \gamma^{*} e^{-} \rightarrow e^{+}\, Z\, Z\, e^{-}$ at the CLIC, with the Z bosons decaying in $e^{+}e^{-}$ and $\mu^{+}\mu^{-}$ pairs, in order to obtain the sensitivities to anomalous quartic gauge couplings. Among these processes, the best bounds on $\frac{a_{0}}{\Lambda^{2}}$ and $\frac{a_{c}}{\Lambda^{2}}$ couplings are obtained from the process $e^{+}e^{-}\rightarrow Z\, Z\, \gamma$, and they are at the order of $10^{-8}$ GeV$^{-2}$. It has been shown that the bounds on anomalous couplings improve up to approximately $10^{-5}$ times for $\frac{a_{0}}{\Lambda^{2}}$ and up to $10^{-6}$ times for $\frac{a_{c}}{\Lambda^{2}}$. In addition, the $e^{+}e^{-} \rightarrow e^{+}\gamma^{*} \gamma^{*} e^{-} \rightarrow e^{+}\, Z\, Z\, e^{-}$ process, due to the number of SM events being negligibly small, provides us an opportunity to examine directly anomalous quartic couplings.
These results show that the CLIC is a suitable platform for probing of anomalous $ZZ\gamma\gamma$ coupling in $e^{+}e^{-}\rightarrow Z\, Z\, \gamma$ as well as $e^{+}e^{-} \rightarrow e^{+}\gamma^{*} e^{-} \rightarrow e^{+} Z\, Z\, e^{-}$ and $e^{+}e^{-} \rightarrow e^{+}\gamma^{*} \gamma^{*} e^{-} \rightarrow e^{+}\, Z\, Z\, e^{-}$ processes.

\pagebreak

\pagebreak

\begin{figure}
\includegraphics[width=0.53\columnwidth]{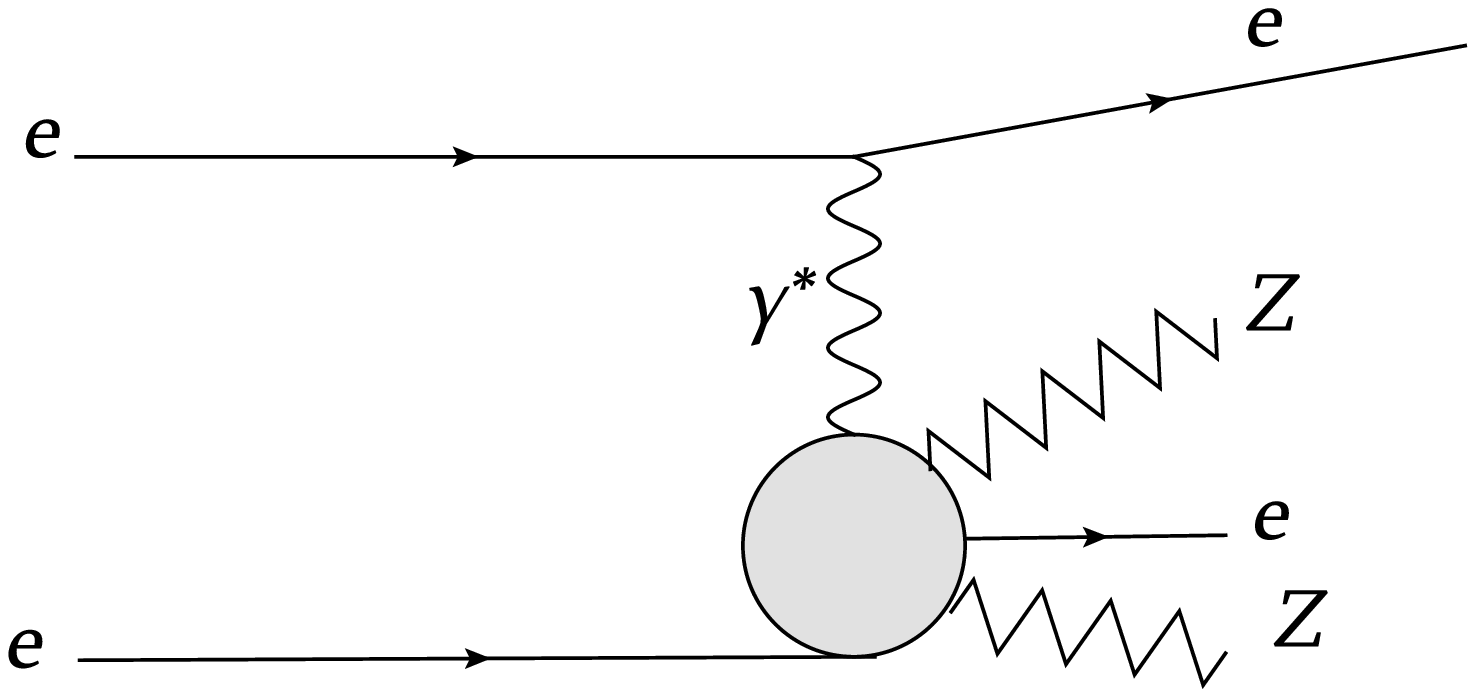}
\caption{Representative diagram for the process $e^{+}e^{-}\rightarrow e^{+}\gamma^{*} e^{-}\rightarrow e^{+}\,Z\,Z\,e^{-}$.
\label{fig1}}
\end{figure}

\begin{figure}
\includegraphics [width=0.5\columnwidth] {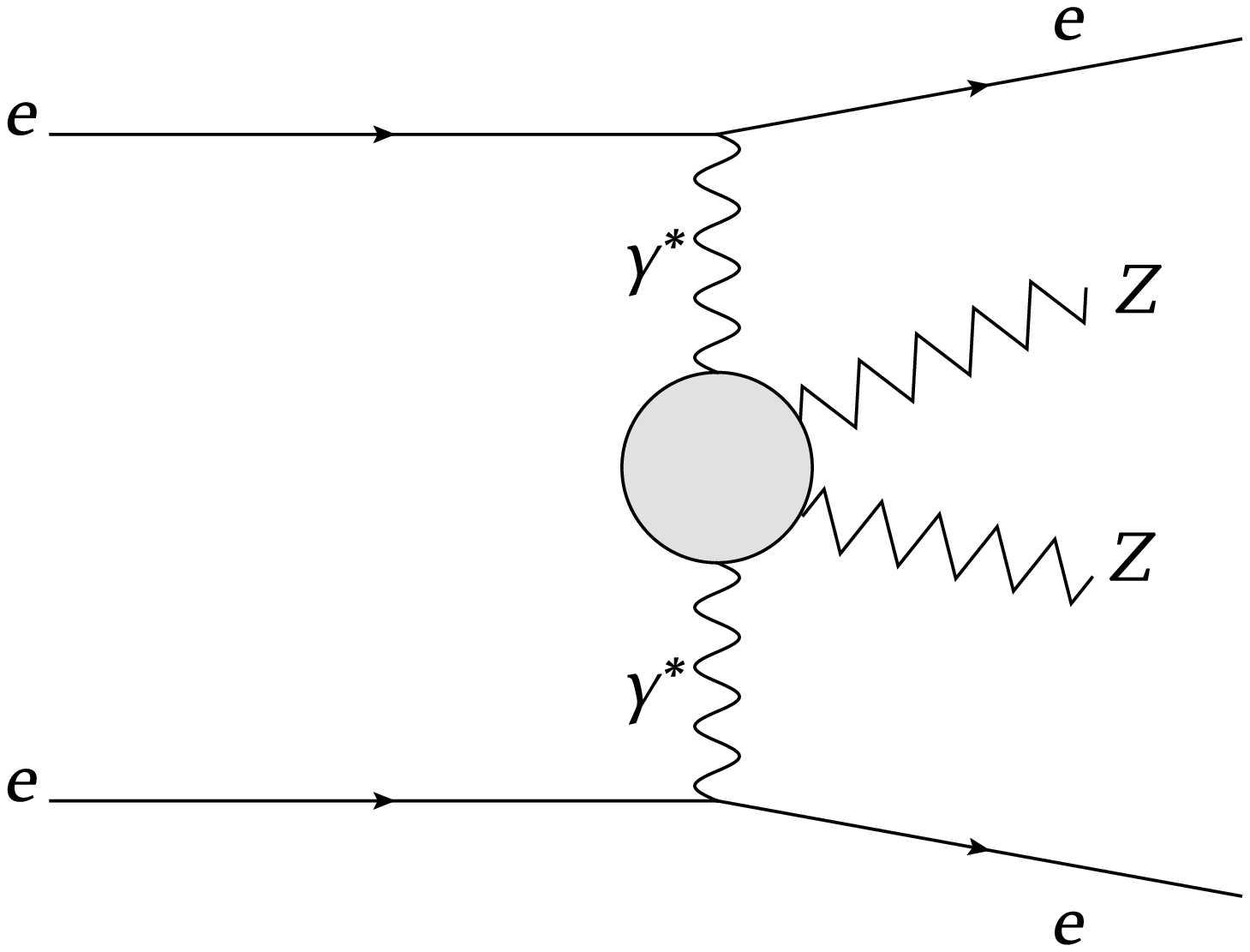}
\caption{Representative diagram for the process $e^{+}e^{-} \rightarrow e^{+}\gamma^{*} \gamma^{*} e^{-} \rightarrow e^{+}\,Z\, Z\,e^{-}$.
\label{fig2}}
\end{figure}

\begin{figure}
\includegraphics [width=0.85\columnwidth] {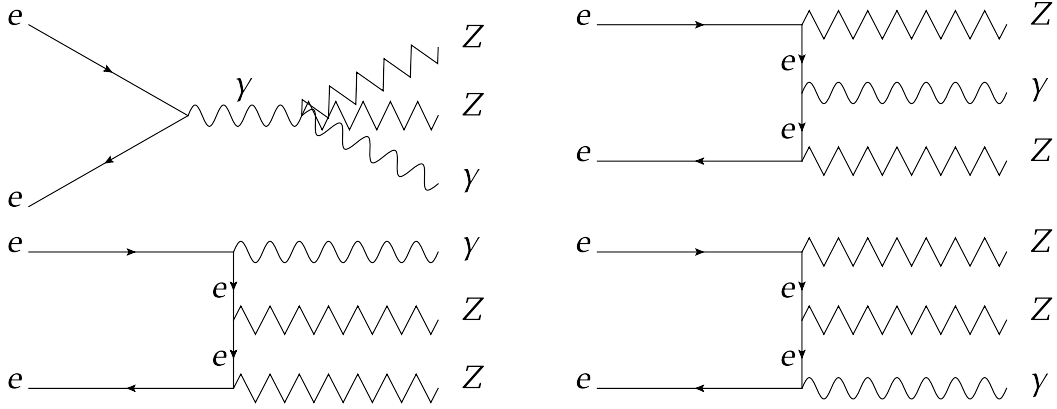}
\caption{Tree-level Feynman diagrams for the process $e^{+}e^{-}\rightarrow Z\, Z\,\gamma$.
\label{fig3}}
\end{figure}

\begin{figure}
\includegraphics [width=0.85\columnwidth] {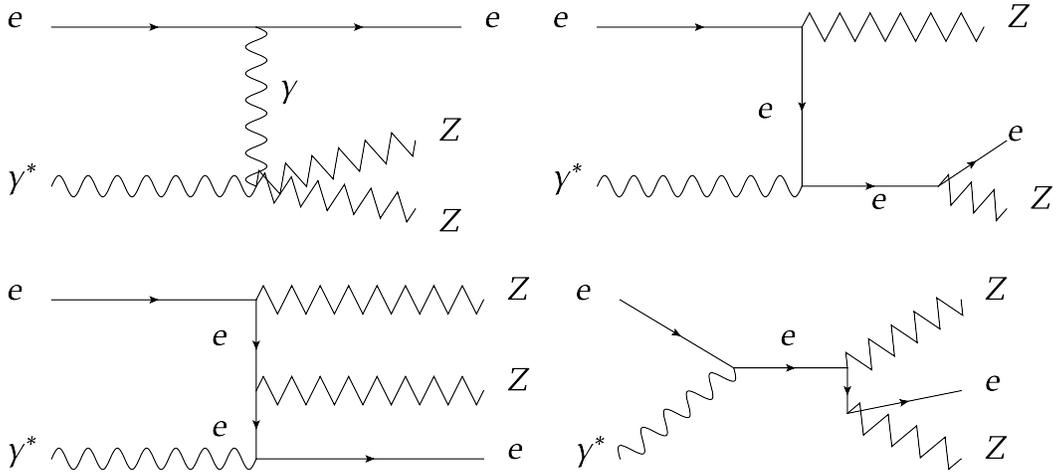}
\caption{Tree-level Feynman diagrams for the subprocess $e^{-}\gamma^{*} \rightarrow Z\,Z\,e^{-}$.
\label{fig4}}
\end{figure}

\begin{figure}
\includegraphics [width=0.4\columnwidth]{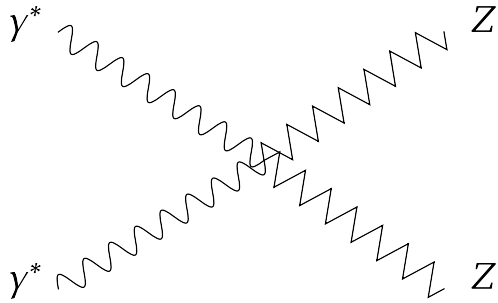}
\caption{Tree-level Feynman diagram for the subprocess $\gamma^{*} \gamma^{*}\rightarrow Z\,Z$.
\label{fig5}}
\end{figure}

\begin{figure}
\includegraphics [width=1.1\columnwidth] {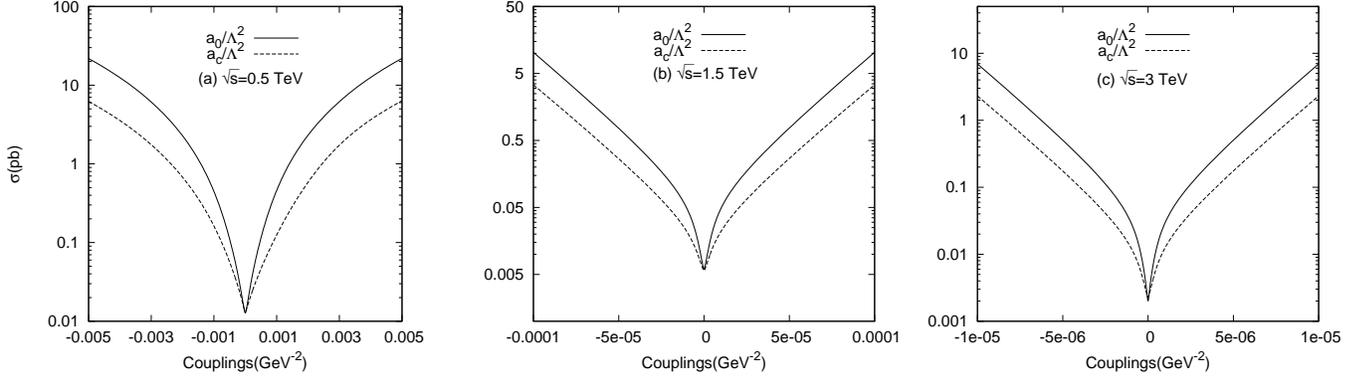}
\caption{The total cross sections as a function of the anomalous $\frac{a_{0}}{\Lambda^{2}}$ and $\frac{a_{c}}{\Lambda^{2}}$ couplings for the process $e^{+}e^{-}\rightarrow Z\, Z\,\gamma$ at the CLIC with $\sqrt{s}=0.5,1.5$ and $3$ TeV.
\label{fig6}}
\end{figure}

\begin{figure}
\includegraphics [width=1.1\columnwidth] {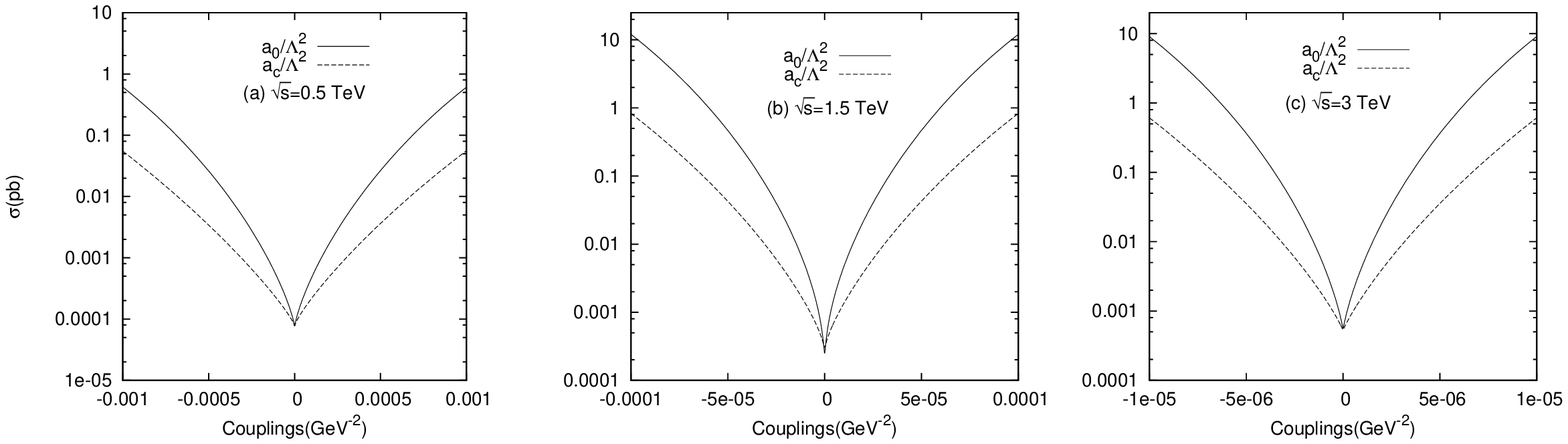}
\caption{The total cross sections as a function of the anomalous $\frac{a_{0}}{\Lambda^{2}}$ and $\frac{a_{c}}{\Lambda^{2}}$ couplings for the process $e^{+}e^{-} \rightarrow e^{+}\gamma^{*} e^{-} \rightarrow e^{+} Z\,Z\,e^{-}$ at the CLIC with $\sqrt{s}=0.5,1.5$ and $3$ TeV.
\label{fig7}}
\end{figure}

\begin{figure}
\includegraphics [width=1.1\columnwidth] {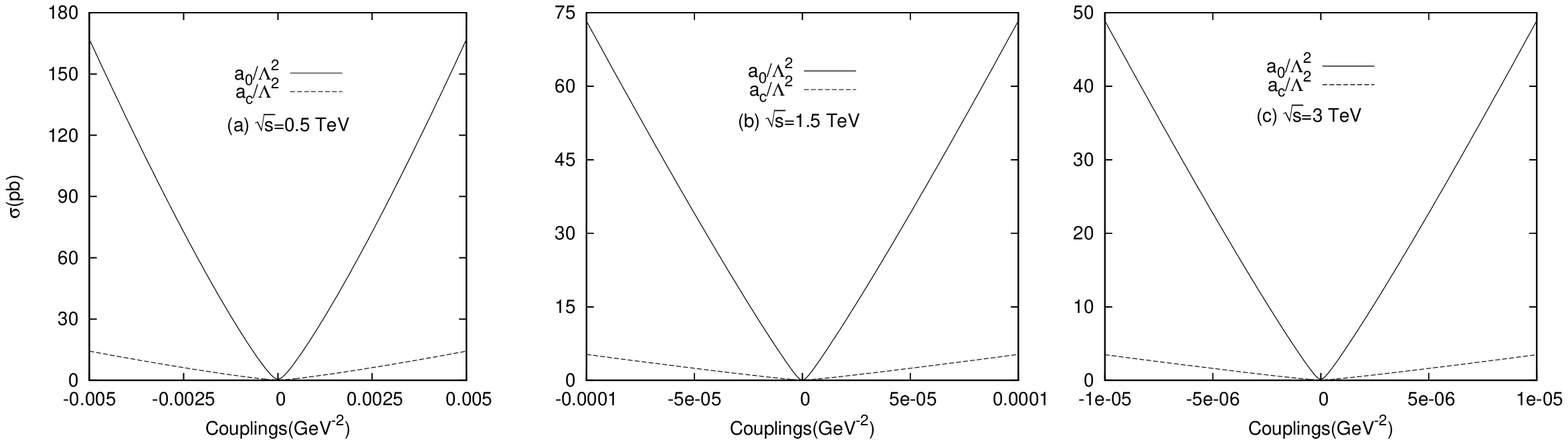}
\caption{The total cross sections as a function of the anomalous $\frac{a_{0}}{\Lambda^{2}}$ and $\frac{a_{c}}{\Lambda^{2}}$ couplings for the process $e^{+}e^{-} \rightarrow e^{+}\gamma^{*} \gamma^{*} e^{-} \rightarrow e^{+}\, Z\, Z\, e^{-}$ at the CLIC with $\sqrt{s}=0.5,1.5$ and $3$ TeV.
\label{fig8}}
\end{figure}

\begin{table}
\caption{The three stages of the CLIC. Here $\sqrt{s}$ is the center-of-mass energy, $L$ is the total luminosity \cite{18}.
\label{tab1}}
\begin{ruledtabular}
\begin{tabular}{ccccc}
Parameter& Unit& Stage\,$1$& Stage\,$2$& Stage\,$3$ \\
\hline
$\sqrt{s}$& TeV& $0.5$& $1.5$& $3$ \\
$L$& fb$^{-1}$& $230$& $320$& $590$  \\
\end{tabular}
\end{ruledtabular}
\end{table}

\begin{table}
\caption{The SM background cross sections for the processes $e^{+}e^{-}\rightarrow Z\, Z\, \gamma$ and
$e^{+}e^{-} \rightarrow e^{+}\gamma^{*} e^{-} \rightarrow e^{+} Z\, Z\, e^{-}$ at the CLIC. Here, we impose the acceptance cuts on the pseudorapidities  $|\eta^{e,\,\gamma}|<2.5$ and the transverse momentums $p_{T}^{\,\gamma}>20 \:$ GeV, $p_{T}^{\,e}>25 \:$ GeV  for electron and photon in the final state of both processes.
\label{tab1}}
\begin{ruledtabular}
\begin{tabular}{ccc}
Process & $\sqrt{s}$ (TeV)& SM Cross Section\,(pb$^{-1}$) \\
\hline
$$ & $0.5$& $1.64\times 10^{-2}$ \\
$e^{+}e^{-}\rightarrow Z\, Z\, \gamma$& $1.5$& $5.69\times 10^{-3}$  \\
$$& $3$& $2.30\times 10^{-3}$  \\
\hline
$$ & $0.5$& $7.68\times 10^{-5}$ \\
$e^{+}e^{-} \rightarrow e^{+}\gamma^{*} e^{-} \rightarrow e^{+} Z\, Z\, e^{-}$& $1.5$& $3.59\times 10^{-4}$  \\
$$& $3$& $5.24\times 10^{-4}$  \\
\hline
\end{tabular}
\end{ruledtabular}
\end{table}

\begin{table}
\caption{The number of observed events $N_{obs}$ and corresponding values for upper limits $N_{up}$ at
$95\%$ C.L. for the process $e^{+}e^{-}\rightarrow Z\, Z\, \gamma$ with different values of luminosity and center-of-mass energy.
\label{tab1}}
\begin{ruledtabular}
\begin{tabular}{cccc}
$\sqrt{s}$ (TeV)& $L_{int}$(fb$^{-1}$)& $N_{obs}$& $N_{up}$ \\
\hline
$0.5$& $10$& $1$& $4.74$ \\
$0.5$& $50$& $4$& $9.15$  \\
$0.5$& $100$& $7$& $13.15$  \\
\hline
$1.5$& $10$& $0$& $3$  \\
$1.5$& $100$& $3$& $7.75$  \\
$1.5$& $200$& $5$& $10.51$  \\
$1.5$& $320$& $8$& $14.43$  \\
\hline
$3$& $10$& $0$& $3$  \\
$3$& $100$& $1$& $4.74$  \\
$3$& $300$& $3$& $7.75$  \\
$3$& $590$& $6$& $11.84$  \\
\end{tabular}
\end{ruledtabular}
\end{table}

\begin{table}
\caption{The number of observed events $N_{obs}$ and associated values for upper limits $N_{up}$ at
$95\%$ C.L. for the process $e^{+}e^{-} \rightarrow e^{+}\gamma^{*} e^{-} \rightarrow e^{+} Z Z e^{-}$ with different values of luminosity and center-of-mass energy.
\label{tab1}}
\begin{ruledtabular}
\begin{tabular}{cccc}
$\sqrt{s}$ (TeV)& $L_{int}$(fb$^{-1}$)& $N_{obs}$& $N_{up}$ \\
\hline
$0.5$& $10$& $0$& $3$ \\
$0.5$& $50$& $0$& $3$  \\
$0.5$& $100$& $0$& $3$  \\
$0.5$& $230$& $0$& $3$ \\
\hline
$1.5$& $10$& $0$& $3$  \\
$1.5$& $100$& $0$& $3$  \\
$1.5$& $200$& $0$& $3$  \\
$1.5$& $320$& $1$& $4.74$  \\
\hline
$3$& $10$& $0$& $3$  \\
$3$& $100$& $0$& $3$  \\
$3$& $300$& $1$& $4.74$  \\
$3$& $590$& $1$& $4.74$  \\
\end{tabular}
\end{ruledtabular}
\end{table}

\begin{table}
\caption{The anomalous quartic gauge coupling parameters $\frac{a_{0}}{\Lambda^{2}}$ and $\frac{a_{c}}{\Lambda^{2}}$ at $95\%$ C.L.
for the process $e^{+}e^{-}\rightarrow Z\, Z\, \gamma$ with various CLIC luminosities. The center of mass
energies of the process are taken to be $\sqrt{s}=0.5,1.5$ and $3$ TeV.
\label{tab2}}
\begin{ruledtabular}
\begin{tabular}{cccc}
$\sqrt{s}$ (TeV)& $L_{int}$(fb$^{-1}$)& $\frac{a_{0}}{\Lambda^{2}}$(GeV$^{-2}$)& $\frac{a_{c}}{\Lambda^{2}}$ (GeV$^{-2}$)\\
\hline
$0.5$& $10$& $[-3.24\times 10^{-4};\, 3.12\times 10^{-4}]$& $[-6.19\times 10^{-4};\, 5.70\times 10^{-4}]$ \\
$0.5$& $50$& $[-1.72\times 10^{-4};\, 1.59\times 10^{-4}]$& $[-3.34\times 10^{-4};\, 2.86\times 10^{-4}]$  \\
$0.5$& $100$& $[-1.27\times 10^{-4};\, 1.14\times 10^{-4}]$& $[-2.53\times 10^{-4};\, 2.02\times 10^{-4}]$\\
$0.5$& $230$& $[-1.01\times 10^{-4};\, 0.88\times 10^{-4}]$& $[-2.01\times 10^{-4};\, 1.53\times 10^{-4}]$ \\
\hline
$1.5$& $10$& $[-7.68\times 10^{-6};\, 7.66\times 10^{-6}]$& $[-1.35\times 10^{-5};\, 1.34\times 10^{-5}]$ \\
$1.5$& $100$& $[-3.34\times 10^{-6};\, 3.30\times 10^{-6}]$& $[-5.87\times 10^{-6};\, 5.75\times 10^{-6}]$  \\
$1.5$& $200$& $[-2.41\times 10^{-6};\, 2.36\times 10^{-6}]$& $[-4.24\times 10^{-6};\, 4.13\times 10^{-6}]$\\
$1.5$& $320$& $[-2.06\times 10^{-6};\, 2.02\times 10^{-6}]$& $[-3.61\times 10^{-6};\, 3.51\times 10^{-6}]$ \\
\hline
$3$& $10$& $[-9.61\times 10^{-8};\, 9.60\times 10^{-8}]$& $[-1.67\times 10^{-7};\, 1.67\times 10^{-7}]$ \\
$3$& $100$& $[-3.56\times 10^{-8};\, 3.55\times 10^{-8}]$& $[-6.17\times 10^{-8};\, 6.14\times 10^{-8}]$  \\
$3$& $300$& $[-2.22\times 10^{-8};\, 2.22\times 10^{-8}]$& $[-3.86\times 10^{-8};\, 3.84\times 10^{-8}]$\\
$3$& $590$& $[-1.76\times 10^{-8};\, 1.75\times 10^{-8}]$& $[-3.06\times 10^{-8};\, 3.04\times 10^{-8}]$ \\
\end{tabular}
\end{ruledtabular}
\end{table}

\begin{table}
\caption{The anomalous quartic gauge coupling parameters $\frac{a_{0}}{\Lambda^{2}}$ and $\frac{a_{c}}{\Lambda^{2}}$ at $95\%$ C.L.
for the process $e^{+}e^{-} \rightarrow e^{+}\gamma^{*} e^{-} \rightarrow e^{+} Z Z e^{-}$ with various CLIC luminosities. The center of mass
energies of the process are taken to be $\sqrt{s}=0.5,1.5$ and $3$ TeV.
\label{tab3}}
\begin{ruledtabular}
\begin{tabular}{cccc}
$\sqrt{s}$ (TeV)& $L_{int}$(fb$^{-1}$)& $\frac{a_{0}}{\Lambda^{2}}$(GeV$^{-2}$)& $\frac{a_{c}}{\Lambda^{2}}$ (GeV$^{-2}$)\\
\hline
$0.5$& $10$& $[-3.30\times 10^{-4};\, 3.29\times 10^{-4}]$& $[-1.10\times 10^{-3};\, 1.09\times 10^{-3}]$ \\
$0.5$& $50$& $[-1.48\times 10^{-4};\, 1.47\times 10^{-4}]$& $[-4.92\times 10^{-4};\, 4.87\times 10^{-4}]$  \\
$0.5$& $100$& $[-1.04\times 10^{-4};\, 1.03\times 10^{-4}]$& $[-3.47\times 10^{-4};\, 3.40\times 10^{-4}]$\\
$0.5$& $230$& $[-0.69\times 10^{-4};\, 0.68\times 10^{-4}]$& $[-2.29\times 10^{-4};\, 2.22\times 10^{-4}]$ \\
\hline
$1.5$& $10$& $[-7.43\times 10^{-6};\, 7.43\times 10^{-6}]$& $[-2.83\times 10^{-5};\, 2.83\times 10^{-5}]$ \\
$1.5$& $100$& $[-2.30\times 10^{-6};\, 2.29\times 10^{-6}]$& $[-8.74\times 10^{-6};\, 8.70\times 10^{-6}]$  \\
$1.5$& $200$& $[-1.58\times 10^{-6};\, 1.57\times 10^{-6}]$& $[-6.02\times 10^{-6};\, 5.96\times 10^{-6}]$\\
$1.5$& $320$& $[-1.20\times 10^{-6};\, 1.19\times 10^{-6}]$& $[-4.59\times 10^{-6};\, 4.54\times 10^{-6}]$ \\
\hline
$3$& $10$& $[-8.52\times 10^{-7};\, 8.52\times 10^{-7}]$& $[-3.28\times 10^{-6};\, 3.28\times 10^{-6}]$ \\
$3$& $100$& $[-2.60\times 10^{-7};\, 2.60\times 10^{-7}]$& $[-9.99\times 10^{-7};\, 9.97\times 10^{-7}]$  \\
$3$& $300$& $[-1.82\times 10^{-7};\, 1.82\times 10^{-7}]$& $[-6.98\times 10^{-7};\, 6.96\times 10^{-7}]$\\
$3$& $590$& $[-1.18\times 10^{-7};\, 1.18\times 10^{-7}]$& $[-4.55\times 10^{-7};\, 4.51\times 10^{-7}]$ \\
\end{tabular}
\end{ruledtabular}
\end{table}

\begin{table}
\caption{The anomalous quartic gauge coupling parameters $\frac{a_{0}}{\Lambda^{2}}$ and $\frac{a_{c}}{\Lambda^{2}}$ at $95\%$ C.L.
for the process $e^{+}e^{-} \rightarrow e^{+}\gamma^{*} \gamma^{*} e^{-} \rightarrow e^{+}\, Z\, Z\, e^{-}$ with various CLIC luminosities. The center of mass
energies of the process are taken to be $\sqrt{s}=0.5,1.5$ and $3$ TeV.
\label{tab2}}
\begin{ruledtabular}
\begin{tabular}{cccc}
$\sqrt{s}$ (TeV)& $L_{int}$(fb$^{-1}$)& $\frac{a_{0}}{\Lambda^{2}}$(GeV$^{-2}$)& $\frac{a_{c}}{\Lambda^{2}}$ (GeV$^{-2}$)\\
\hline
$0.5$& $10$& $[-1.00\times 10^{-4};\, 1.00\times 10^{-4}]$& $[-3.41\times 10^{-4};\, 3.41\times 10^{-4}]$ \\
$0.5$& $50$& $[-0.45\times 10^{-4};\, 0.45\times 10^{-4}]$& $[-1.52\times 10^{-4};\, 1.52\times 10^{-4}]$  \\
$0.5$& $100$& $[-0.32\times 10^{-4};\, 0.32\times 10^{-4}]$& $[-1.08\times 10^{-4};\, 1.08\times 10^{-4}]$\\
$0.5$& $230$& $[-0.21\times 10^{-4};\, 0.21\times 10^{-4}]$& $[-0.71\times 10^{-4};\, 0.71\times 10^{-4}]$ \\
\hline
$1.5$& $10$& $[-3.01\times 10^{-6};\, 3.01\times 10^{-6}]$& $[-1.12\times 10^{-5};\, 1.12\times 10^{-5}]$ \\
$1.5$& $100$& $[-9.51\times 10^{-7};\, 9.51\times 10^{-7}]$& $[-3.54\times 10^{-6};\, 3.54\times 10^{-6}]$  \\
$1.5$& $200$& $[-6.73\times 10^{-7};\, 6.73\times 10^{-7}]$& $[-2.51\times 10^{-6};\, 2.51\times 10^{-6}]$\\
$1.5$& $320$& $[-5.32\times 10^{-7};\, 5.32\times 10^{-7}]$& $[-1.98\times 10^{-6};\, 1.98\times 10^{-6}]$ \\
\hline
$3$& $10$& $[-3.68\times 10^{-7};\, 3.68\times 10^{-7}]$& $[-1.38\times 10^{-6};\, 1.38\times 10^{-6}]$ \\
$3$& $100$& $[-1.17\times 10^{-7};\, 1.17\times 10^{-7}]$& $[-4.36\times 10^{-7};\, 4.36\times 10^{-7}]$  \\
$3$& $300$& $[-6.72\times 10^{-8};\, 6.72\times 10^{-8}]$& $[-2.52\times 10^{-7};\, 2.52\times 10^{-7}]$\\
$3$& $590$& $[-4.79\times 10^{-8};\, 4.79\times 10^{-8}]$& $[-1.79\times 10^{-7};\, 1.79\times 10^{-7}]$ \\
\end{tabular}
\end{ruledtabular}
\end{table}

\end{document}